\documentclass[aps,prev,twocolumn,preprintnumbers,floatfix,nofootinbib]{revtex4-1}
\usepackage{graphicx}
\usepackage{bm}
\usepackage{times}
\usepackage{slashed}
\usepackage{color}
\usepackage{amsmath}
\usepackage{amsthm}
\usepackage{subfigure}
\usepackage{hyperref}
\hypersetup{
    colorlinks=true,
    linkcolor=blue,
    filecolor=magenta,      
    urlcolor=blue,
    citecolor=blue,
    pdftitle={Sharelatex Example},
    pdfpagemode=FullScreen
    }
\newcommand{\gmu}{$g-2$}
\newcommand{\bea}{\begin{eqnarray}}
\newcommand{\eea}{\end{eqnarray}}

\newcommand{\e}{Eq.}

\begin{document}

\title{Dead or Alive? Implications of the Muon Anomalous Magnetic Moment for 3-3-1 Models}

\author{Álvaro S. de Jesus$^{a}$}
\email{alvarosdj@ufrn.edu.br}

\author{Sergey Kovalenko$^{b}$}
\email{sergey.kovalenko@unab.cl}

\author{C. A. de S. Pires$^{c}$}
\email{cpires@fisica.ufpb.br}

\author{Farinaldo S. Queiroz$^{a}$}
\email{farinaldo.queiroz@iip.ufrn.br}

\author{Yoxara S. Villamizar$^{a}$}
\email{yoxara@ufrn.edu.br}

\affiliation{$^a$International Institute of Physics, Universidade Federal do Rio Grande do Norte,
Campus Universit\'ario, Lagoa Nova, Natal-RN 59078-970, Brazil\\
$^b$ Departamento de Ciencias F\'isicas, 
Universidad Andres Bello, Sazi\'e 2212, Santiago, Chile\\
$^c$Departamento de F\'isica, Universidade Federal da Para\'iba,
Caixa Postal 5008, 58051-970, Joao Pessoa, PB, Brazil }

\begin{abstract}
We have witnessed a persistent puzzling anomaly in the muon magnetic moment that cannot be accounted for in the Standard Model even considering the existing large hadronic uncertainties. A new measurement is forthcoming, and it might give rise to a $5\sigma$ claim for physics beyond the Standard Model. Motivated by it, we explore the implications of this new result to five models based on the $SU(3)_C \times SU(3)_L \times U(1)_N$ gauge symmetry and put our conclusions into perspective with LHC bounds. We show that previous conclusions found in the context of such models change if there are more than one heavy particle running in the loop. Moreover, having in mind the projected precision aimed by the g-2 experiment at FERMILAB, we place lower mass bounds on the particles that contribute to muon anomalous magnetic moment assuming the anomaly is otherwise resolved. Lastly, we discuss how these models could accommodate such anomaly in agreement with current bounds. 
\end{abstract}


\maketitle

\section{Introduction}
\label{intro}

The Standard Model (SM) offers an excellent description of the strong and electroweak interactions in nature. Its theoretical predictions, calculated beyond tree level, are compatible with experimental measurements with unprecedented accuracy. Nevertheless, there are many open questions, which cannot be addressed within the SM.  The long-standing muon anomalous magnetic moment anomaly, if experimentally confirmed, is one of them.  The deviation of the magnetic moment of any charged fermion from its Dirac prediction, $g/2=1$, is quantified by the aforementioned anomalous magnetic moment $(g-2)/2$. For the case of the muon, the SM predicts,

\begin{equation}
    a_{\mu}^{SM}=\frac{g-2}{2}= 116591802 (2)(42)(26) \times 10^{-11}.
    \label{amuSM}
\end{equation}This value takes into account the electromagnetic, weak, and hadronic corrections (See \cite{Lindner:2016bgg} for a recent review). The theoretical error showed in Eq. \eqref{amuSM} corresponds to the electroweak, lowest-order hadronic, and higher-order hadronic contributions, respectively. Several diagrams that contribute to $a^{SM}_\mu$ have been computed beyond one-loop. 

Experiments have measured the muon anomalous magnetic moment using the principle of Larmor precession, whose frequency is proportional to the magnetic field which the charged particle is subject to. As the theoretical and experimental errors shrank, a discrepancy, quantified by  $\Delta a_\mu= a_\mu^{exp}-a_\mu^{SM}$, was observed between the Standard Model prediction and the experimental measurements. Comparing the SM prediction with the measurements from Brookhaven National Lab \cite{Bennett:2002jb,Bennett:2006fi}, we get,

\begin{equation*}
    \Delta a_\mu= (261 \pm 78)\times 10^{-11} \quad (3.3\sigma)
\end{equation*}  \begin{center} \cite{Prades:2009tw,Tanabashi:2018oca} - (2009)\end{center}
\begin{equation*}
    \Delta a_\mu=  (325 \pm 80)\times 10^{-11} \quad (4.05\sigma)
\end{equation*} \begin{center}\cite{Benayoun:2012wc} - (2012)\end{center}
\begin{equation*}
     \Delta a_\mu= (287 \pm 80)\times 10^{-11} \quad (3.6\sigma)
\end{equation*}
\begin{center}
    \cite{Blum:2013xva} - (2013)
\end{center}
\begin{equation*}
    \Delta a_\mu=  (377 \pm 75)\times 10^{-11}\quad (5.02\sigma)
\end{equation*}
\begin{center}
    \cite{Benayoun:2015gxa} - (2015)
\end{center}
\begin{equation*}
    \Delta a_\mu=  (313 \pm 77)\times 10^{-11} \quad (4.1\sigma)
\end{equation*}
\begin{center}
    \cite{Jegerlehner:2017lbd}- (2017)
\end{center}
\begin{equation}
     \Delta a_\mu=  (270 \pm 36)\times 10^{-11} \quad (3.7\sigma) \label{deltasigmavalues}
\end{equation}
\begin{center}
    \cite{Keshavarzi:2018mgv} - (2018)
\end{center}

The different values quoted in Eq.\eqref {deltasigmavalues} refer to different studies where the overall SM contribution was reassessed based on different calculations of the hadronic contribution. According to the Particle Data Group (PDG), the current discrepancy reads $3.3\sigma$, but the PDG review already acknowledges recent studies where the significance approaches $4\sigma$. It is clear that such hadronic contribution blurs the significance of this anomaly, but it appears a sign of new physics. We will be conservative in our study and adopt $\Delta a_\mu=(261 \pm 78)\times 10^{-11}$ \cite{Tanabashi:2018oca}.  

Fortunately, there are two experiments (g-2 at FERMILAB \cite{Grange:2015fou} and Muon g-2 at J-PARC \cite{Abe:2019thb}) that will be able to push down the error bar and increase the discrepancy if the central value remains the same. The goal is to bring the error down by a factor of four. In particular, the g-2 experiment is about to announce new results. Keeping the central value of the previous measurement, the g-2 experiment has the potential to claim the first $5\sigma$ signal after the Higgs boson discovery, and with the expected theoretical improvements \cite{Campanario:2019mjh,Davier:2019can} this significance could increase to nearly $8\sigma$ \cite{Kronfeld:2013uoa}. Such observation will have profound implications for particle physics. If the anomaly is not confirmed though, we will place conservative $1\sigma$ bounds by enforcing the overall contribution of a given model to g-2 to be at most $78 \times 10^{-11}$. Furthermore, considering the sensitivity aimed by the g-2 collaboration and the expected reduction of the theoretical uncertainties of the hadronic contribution \cite{Carey:2009zzb}, we can further impose a projected $1\sigma$ bound by requiring  $\Delta a_\mu < 34 \times 10^{-11}$ \cite{Carey:2009zzb}. In summary, we are arguably on the climax of the muon magnetic moment history, and for this reason, our investigation is timely important.

Instead of exploring g-2 in a simplified model where there are few new physics contributions to g-2 and the experimental constraints are easier to bypass, we investigate g-2 in Ultra-Violet complete models based on the $SU(3)_C \times SU(3)_L \times U(1)_X$ (3-3-1) gauge symmetry. Such models are well-motivated for several reasons. They are capable of addressing the number of fermion generations due to anomaly cancellations and QCD asymptotic freedom requirements. Such models are anomaly free if there is an equal number of triplets and antitriplets fermion multiplets. The anomaly cancellation does not occur generation by generation, it does when all three fermion generations are considered. In this way, we need three fermion generations to have an anomaly free model. On the other side, asymptotic freedom of QCD requires less than 17 quarks.  Thus, models with three fermion generations stand as the simplest non-trivial anomaly free representation of the $SU(3)_L \times U(1)_N$ gauge group \cite{Foot:1992rh}.

These models can explain neutrino masses \cite{Montero:2000rh, Tully:2000kk,Montero:2001ts,Cortez:2005cp,Cogollo:2009yi,Cogollo:2010jw,Cogollo:2008zc,Okada:2015bxa,Vien:2018otl,carcamoHernandez:2018iel,Nguyen:2018rlb,Pires:2018kaj,CarcamoHernandez:2019iwh,CarcamoHernandez:2019vih,CarcamoHernandez:2020pnh} and dark matter \cite{Fregolente:2002nx,Hoang:2003vj,deS.Pires:2007gi,Mizukoshi:2010ky,Profumo:2013sca,Dong:2013ioa,Dong:2013wca,Cogollo:2014jia,Dong:2014wsa,Dong:2014esa,Carvajal:2017gjj,Montero:2017yvy,Huong:2019vej}. Due to the enlarged scalar sector they are entitled to a rich phenomenology that can be explored in the context of meson oscillations, colliders, flavor violation \cite{Cabarcas:2012uf,Hue:2017lak}, among others \cite{Montero:2011tg,Santos:2017jbv,Barreto:2017xix,DeConto:2017ebj}. 

That said, there are models based on this gauge symmetry that have become quite popular. They are known as {\it Minimal 3-3-1}  \cite{Pisano:1991ee}, 3-3-1 with right-handed neutrinos  ({\it 3-3-1 r.h.n}), \cite{Hoang:1996gi,Hoang:1995vq}, 3-3-1 with neutral lepton ({\it 3-3-1 LHN}) \cite{Mizukoshi:2010ky,Catano:2012kw},  {\it Economical 3-3-1} \cite{Dong:2006mg,Dong:2008ya,Martinez:2014lta}, and {\it 3-3-1 with exotic leptons} \cite{Ponce:2001jn,Ponce:2002fv,Anderson:2005ab,Cabarcas:2013jba}. Each model gives rise to several new contributions to $a_\mu$, and these corrections come either from gauge bosons, or scalar fields, or new fermions. Some of them induce a negative contribution to $a_\mu$. In the past, these corrections to $a_\mu$ have been studied as a function of the masses of the particles. The conclusions can be misleading when there are multiple particles contributing to g-2. Thus, we believe that the proper way to present results in the context of 3-3-1 models is by presenting results in terms of the energy scale of symmetry breaking. The individual contributions to g-2 are not particularly relevant, but the overall correction to g-2, because there might be cancellations. The main corrections to g-2 stem from new gauge bosons whose masses are determined by the energy scale at which the 3-3-1 symmetry is broken. Therefore, we can write down all individual contributions in terms of this energy scale and later sum them up to derive the overall contribution to g-2. We do this exercise for the five models under study. Moreover, in this way, we can connect our findings to existing bounds rising from collider physics. We believe that our findings will represent a new direction concerning 3-3-1 model building endeavors if the g-2 anomaly is confirmed. 

There were studies of the muon magnetic moment in the context of 3-3-1 models in the past \cite{Ky:2000ku,Kelso:2013zfa,Binh:2015cba,Cogollo:2017foz,DeConto:2016ith,CarcamoHernandez:2020pxw}. We extend them by investigating five different models simultaneously and putting them into perspective with existing bounds. Moreover, we show that previous conclusions found in the literature are not valid in the presence of two heavy new fields in the loop. At loop level, there can be more than one heavy particle in the loop that contributes to the muon magnetic moment. This is a common feature in 3-3-1 models as they have a rich particle spectrum. We show how our conclusions change depending on the masses of those particles and draw robust conclusions. We provide a Mathematica notebook, available at \cite{numericalcode}.

\section{3-3-1 Models}

It is important to highlight some key features of the models based on the $SU(3)_C \times SU(3)_L \times \times U(1)_N$ gauge symmetry before discussing each model individually. Enlarging the $SU(2)_L$ symmetry to $SU(3)_L$ implies that the fermion generations are now triplets or antitriplets under $SU(3)_L$. After the $SU(3)_L \times U(1)_N$ symmetry is spontaneously broken, a remnant $SU(2)_L \times U(1)_Y$ is observed \cite{Borges:2016nne}.  The fermionic and bosonic contents are dictated by the electric charge operator which is generally written as a combination of the diagonal generators of the group as follows, 

\begin{equation}
\frac{Q}{e}=\frac{1}{2} (\lambda_3+ \alpha \lambda_8)+ N I = 
\begin{pmatrix}
     1/2 (1+\frac{\alpha}{\sqrt{3}}) +N  \\
    1/2 (-1+ \frac{\alpha}{\sqrt{3}}) +N \\
     - \frac{ \alpha}{\sqrt{3}} +N
\end{pmatrix},
\end{equation}where $\lambda_{3,8}$ and $I$ are the generators of $SU(3)_L$ and $U(1)_N$, respectively.

As we want to reproduce the Standard Model spectrum, the first two components of the triplet should be a neutrino and a charged lepton. From this requirement, we get $\alpha / \sqrt{3}=-(2N+1)$, which implies that the third component should have a $3N+1$ quantum number under $U(1)_N$. For instance, taking $N=-1/3$, the third component would {\bf be} a right-handed neutrino $\nu_R^c$, or  simply a neutral fermion, $N$. These choices lead to the {\it 3-3-1 r.h.n}., and 3-3-1 LHN. If we took N=0, i.e. $\alpha= -\sqrt{3}$, then the third component would be a positively charged lepton, either $l^c$, or $E$, where $l^c$ is simply the charge-conjugate of the Standard Model lepton $l$, while $E$ is an exotic charged lepton. The last two choices lead to the {\it Minimal 3-3-1} model and {\it 3-3-1 model with exotic leptons}, respectively. These 3-3-1 models usually feature three scalar triplets in the scalar sector, but if there are only two, then, it is called {\it Economical 3-3-1}. There are other possible ways to extend these models by introducing extra singlet fermions under $SU(3)_L$, additional scalar multiplets, etc. The initial motivation behind all these models is the possibility to solve the number of fermion generations. Later on, it was realized that many of them are also capable of explaining neutrino masses, dark matter, and some flavor anomalies \cite{Wei:2017ago}. 

We now move on to a brief description of each model. We emphasize that we will not discuss the models in detail, we will focus rather on the aspects that are relevant to our phenomenology.

\section{MINIMAL 3-3-1}
\label{minimal}

In the {\it Minimal 3-3-1} model the leptotonic triplet is arranged as \cite{Pisano:1991ee,Montero:1992jk},
\begin{equation} f^{a}_L =
\begin{pmatrix}
   \nu^a  \\
    l^a  \\
    (l^c)^a
\end{pmatrix}
\end{equation} where the $a=1,2,3$ is the generation index. As explained above this model is a consequence of taking $\alpha=-\sqrt{3}$. It is well-known that an $SU(3)_L \times U(1)_N$ gauge has 9 gauge bosons. Four of them are identified as the $W^{\pm}$, $Z$ and the photon. There are other five massive gauge bosons known as $W^{\prime \pm}$, $U^{\pm \pm }$ and $Z^\prime$. The $W^\prime$ gauge boson experiences a charge current similar to the W boson of the Standard Model. The $U^{\pm \pm }$ is a doubly charged gauge boson. The dynamics of the gauge boson interactions with leptons are governed by the charged (${\cal L}^{CC}$) and neutral (${\cal L}^{NC}$) currents Lagrangian \cite{Pisano:1991ee},
\begin{equation}
\label{doublyminimal}
{\cal L}^{CC}_l\supset-\frac{g}{2\sqrt{2}}\left[\bar{\nu}\gamma^\mu (1-\gamma_5) C\bar{l}^{T}W^{\prime-}_\mu-\bar{l}\gamma^\mu\gamma_5 C\bar{l}^T
U^{--}_\mu\right],
\end{equation}
\begin{equation}{\cal L}^{NC} \supset
\bar{f}\, \gamma^{\mu} [g_{V}(f) + g_{A}(f)\gamma_5]\, f\,
Z'_{\mu}, \label{ncm}
\end{equation}
where $g_{V}$ and $g_{A}$ are the vector and axial coupling constants, which for charged leptons read, 
\bea g_{A}(l) & &= \frac{g}{2c_W}
\frac{\sqrt{3}\sqrt{1 - 4 s_W^2}}{6},\
g_{V}(l) = 3 g_{A}(l),
\label{hsm}
\eea
with \[g'=g\frac{s_W}{\sqrt{1-\frac{4}{3}s_W^2}},\] where,
$s_W=sen(\theta_W)$, $c_W=cos(\theta_W)$, and $\theta_W$ is the Weinberg angle. All these gauge bosons will contribute to $a_\mu$. We left out the hadronic sector as it is irrelevant to our g-2 study.

The masses of all  fermions and gauge bosons are obtained via the introduction of three scalar triplets and one scalar sextet,

\begin{equation}
    \chi=\begin{pmatrix}
         \chi^-\\
         \chi^{--}\\
         \chi^{0}
    \end{pmatrix},
         \rho=\begin{pmatrix}
         \rho^+\\
         \rho^0\\
         \rho^{++}
    \end{pmatrix},
    \eta=\begin{pmatrix}
         \eta^0\\
         \eta_1^+\\
         \eta_2^+
    \end{pmatrix},
\end{equation}

\begin{equation}
S  =  \left( \begin{array}{ccc} 
\sigma_1^0 & h_2^- & h_1^+\\ 
h_2^- & H_1^{--} & \sigma_2^0\\ 
h_1^+ & \sigma_2^0 & H_1^{++} 
\label{tripletscalars} 
\end{array}  \right), 
\end{equation} where the vacuum expectation value ($vev$) for every one of the neutral components of scalars are $\left<\eta^0\right>=v_\eta$, $\left<\rho^0\right>=v_\rho$, $\left<\chi^0\right>=v_\chi$, $\left<\sigma^0_2\right>=v_{\sigma_2}$ and  $\left<\sigma^0_1\right>=v_{\sigma_1}$. One may notice that after spontaneous symmetry breaking mechanism scalar sextet breaks down to a scalar triplet, doublet and a singlet field \cite{Montero:2000rh}. This scalar sextet is important to generate neutrino masses via a type II seesaw mechanism \cite{Montero:2000rh, Tully:2000kk,Queiroz:2010rj,Ferreira:2019qpf}. Similarly to what occurs in the type II seesaw mechanism, the combination    $v_\eta^2+v_\rho^2+v_{\sigma_2}^2 + 2 v^2_{\sigma_1}$ contributes to the $W$ mass, this allows us to recognize $v_\eta^2+v_\rho^2+v_{\sigma_2}^2 =v^2$, where $v\approx 246$GeV. In this work we consider $v_{\sigma_1}$ to be sufficiently small as required by the $\rho$ parameter \cite{Camargo:2018uzw}. In this way we can take $v_\eta=v_\rho=v_{\sigma_2} \approx v/\sqrt{3}$. We highlight that we did not include the scalar sextet in our calculations of $a_\mu$. The scalar sextet gives rise to a negative contribution to g-2 and small when compared to the doubly charged gauge boson. Therefore, its inclusion in our discussion is simply a matter of completeness.

As a result, the masses of the new gauge bosons are given by \cite{Pisano:1991ee},
\begin{eqnarray}
M_{W^{\prime}}^2 = \frac{g^2}{4}\left( v_{\eta}^2 + v_{\chi}^2 +
v_{\sigma_2}^2 +2v^2_{\sigma_1} \right), \nonumber\\
M_U^2 =\frac{g^2}{4}\left( v_{\rho}^2 + v_{\chi}^2
+ 4 v_{\sigma_2}^2 \right),\nonumber\\
M_{Z^{\prime}}^2 \approx \left( \frac{g^2+\frac{g^{\prime 2}}{3}}{3} \right) v_{\chi}^2.
\label{massWZprime}
\end{eqnarray}

\begin{figure*}[!t]
\centering
\subfigure[\label{feynman1}]{\includegraphics[scale=0.4]{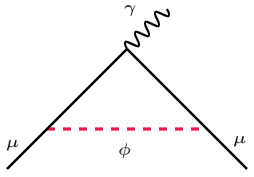}}
\subfigure[\label{feynman2}]{\includegraphics[scale=0.4]{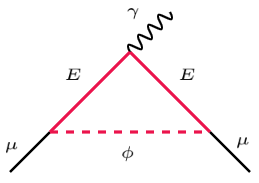}}
\subfigure[\label{feynman3}]{\includegraphics[scale=0.4]{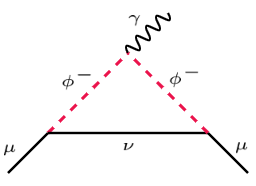}}
\subfigure[\label{feynman4}]{\includegraphics[scale=0.4]{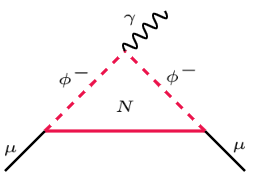}}
\subfigure[\label{feynman5}]{\includegraphics[scale=0.4]{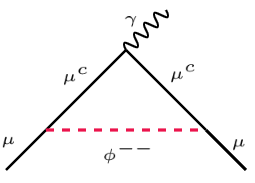}}
\subfigure[\label{feynman6}]{\includegraphics[scale=0.4]{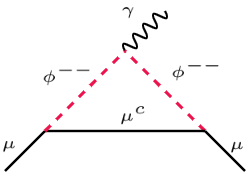}}
\subfigure[\label{feynman7}]{\includegraphics[scale=0.4]{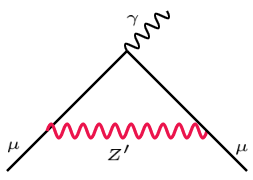}}
\subfigure[\label{feynman8}]{\includegraphics[scale=0.4]{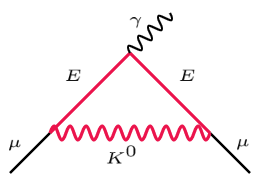}}
\subfigure[\label{feynman9}]{\includegraphics[scale=0.4]{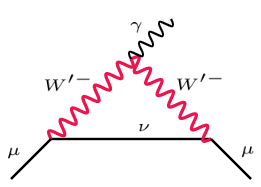}}
\subfigure[\label{feynman10}]{\includegraphics[scale=0.4]{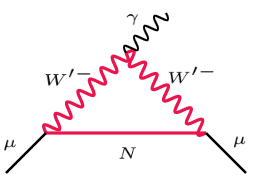}}
\subfigure[\label{feynman11}]{\includegraphics[scale=0.4]{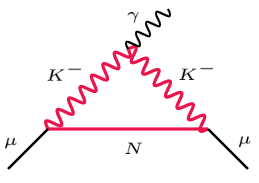}}
\subfigure[\label{feynman12}]{\includegraphics[scale=0.4]{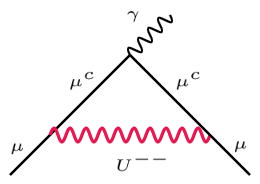}}
\subfigure[\label{feynman13}]{\includegraphics[scale=0.4]{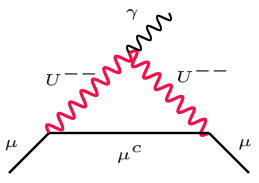}}
\caption{Feynmann diagrams that contribute to the muon anomalous magnetic moment in the 3-3-1 models investigated in this work.}
\end{figure*}

It is important to mention that the relevant interactions involving scalar fields as far as the muon magnetic moment is concerned are, \cite{Foot:1992rh},
\begin{equation}{\cal L} \supset G_l\,\left[ \overline{l_R}\, \nu_L \eta_1^- + \overline{l_R^c}\, \nu_L h_1^+ + \overline{l_R}\, \nu_L h_2^+ + \overline{l_R} l_L R_{\sigma_2} \right]
+ h.c, 
\label{chargedscal}
\end{equation}
 with $G_l=\frac{m_l\sqrt{2}}{2 v_\eta}$ and $R_{\sigma_2}$ the real component of the $\sigma_2^0$, where the masses for $\eta^+_1$, $h^+_1$, $h^+_2$ and $R_{\sigma_2}$ are given by \cite{Tonasse:1996cx},
\bea
\label{Sm}
 M_{\eta_1^+}^2 & \sim & f v_{\chi},\nonumber\\ 
M_{h_1^+,h_2^+} & & \sim v_{\chi} ,\nonumber\\
 M_{R_{\sigma_2}} & &\sim v_{\chi},
\eea where  $f$ is an energy  parameter whose value must lie around $v_{\chi}$\cite{Tonasse:1996cx}. These scalars interact with leptons through the Yukawa Lagrangian in Eq.\eqref{chargedscal} meaning that they couple to leptons proportionally to their masses. Hence, their contribution to $a_\mu$ will be suppressed. The neutral scalar contribution arises via the Feynman diagram in FIG. \ref{feynman1}, the charged scalar via FIG. \ref{feynman3}, the doubly charged scalar through FIGS. \ref{feynman5}-\ref{feynman6}, the $Z^\prime$ via FIG. \ref{feynman7}, the $W^\prime$ via FIG. \ref{feynman9} and the $U^{\pm \pm}$ through FIG. \ref{feynman12}-\ref{feynman13}. The doubly charged gauge boson contribution gives rise to the largest modification of $a_\mu$. If we had included the scalar sextet in our calculations our conclusions would not have changed.

\section{3-3-1 r.h.n}
\label{rhn}
The 3-3-1 model that features right-handed neutrinos is called ({\it 3-3-1 r.h.n}), the third component of the leptonic triplet is replaced by a right-handed neutrino. For this reason, the model has the following leptonic sector \cite{Hoang:1995vq,Hoang:1996gi},

\begin{equation} f^{a}_L =
\begin{pmatrix}
   \nu^a  \\
    l^a  \\
    (\nu^c)^a
\end{pmatrix} ;\,\, l^a_R.
\end{equation}

In this case the five new gauge bosons are the $W^{\prime \pm}$, $Z^{\prime}$, $X^0$ and $X^{0 \dagger}$. As the third component of the fermion triplet is a neutral fermion, there is no doubly charged boson and the $X^0$ and $X^{0 \dagger}$ bosons are neutral. Following the notation in Eq.\eqref{ncm}, the vector and axial-vector couplings of the neutral current are found to be \cite{Hoang:1995vq,Hoang:1996gi},
\begin{equation}
g'_{V}(l) = \frac{g}{4 c_W} \frac{(1 -
4 s_W^2)}{\sqrt{3-4s_W^2}},\
g'_{A}(l) = -\frac{g}{4 c_W \sqrt{3-4s_W^2}},
\label{hsr}
\end{equation}while the charged current takes the form,

\bea
{\cal L} \supset - \frac{g}{2\sqrt{2}}\left[
\overline{\nu^c_R}\, \gamma^\mu (1- \gamma_5) \bar{l}\, W^{\prime -}_\mu  \right],
\label{Irhn}
\eea which is similar to the {\it Minimal 3-3-1} model. It is important to highlight that the $X^0$ and $X^{0\dagger}$ gauge bosons do not contribute to the neutral current.

The scalar sector of the model also features three scalar triplet as follows,

\begin{equation}
    \chi=\begin{pmatrix}
         \chi^0\\
         \chi^{-}\\
         \chi^{0\prime}
    \end{pmatrix},
         \rho=\begin{pmatrix}
         \rho^+\\
         \rho^0\\
         \rho^{+\prime}
    \end{pmatrix},
    \eta=\begin{pmatrix}
         \eta^0\\
         \eta^-\\
         \eta^{0\prime}
    \end{pmatrix}.
    \label{tripletscalars2}
\end{equation}

The spontaneous symmetry breaking mechanism leads to the gauge boson masses,

\bea
M_{Z^{\prime}}^2 = \frac{g^2}{4(3-4s_w^2)}\left( 4 c_W^2 v_{\chi}^2 + \frac{v_{\rho}^2}{c_W^2}
+ \frac{v_{\eta}^2(1-2s_W^2)^2}{c_W^2} \right),
\label{zprimemassmin}\nonumber\\
\eea and

\bea
M_{W^{\prime}}^2 =M_{X^0}^2 = \frac{g^2}{4}\left( v_{\eta}^2 + v_{\chi}^2 \right).
\label{wprimemassmin}
\eea

For simplicity we take $\left<\chi^{0\prime}\right>=v_{\chi}$, $\left< \rho^0\right>=v_\rho$, $\left<\eta^0\right>=v_\eta$, with the other neutral scalar not developing a $vev$ different from zero.  We also assume that the $SU(3)_L \times U(1)_N$ symmetry breaking energy scale, $v_{\chi}$, occurs at energies much higher than the weak scale.  As before the condition $v_\eta^2+v_\rho^2 = v^2$ is obeyed with $v \sim 246 \,\mbox{GeV}$.

The Yukawa Lagrangian involving the charged scalars is essentially the same as in Eq.\eqref{chargedscal} and for the same reason the contributions to $a_\mu$ stemming from charged scalars are small. As far as the scalar sector is concerned, there is a small difference which is due to the presence of a neutral scalar that interacts with the muon through the Yukawa Lagrangian,

\begin{equation}
{\cal L} \supset G_{ab} \bar{f_a}_L \rho e_{b_R},
\end{equation}which leads to 
\begin{equation}
{\cal L} \supset G_s \bar{\mu}\, \mu S_2,
\label{neutralSca}
\end{equation}where $G_s=m_{\mu} \sqrt{2}/(2 v)$. 

As the scalar couples to muons proportionally to the muon mass, we conclude that their corrections to  $a_\mu$ are dwindled. We included all these contributions in our computations anyway. The masses of these scalars are given by, 
\bea 
 M^{2}_{S_{2}} & = & \frac{1}{2}(v_{\chi}^{2}+2v^{2}(2\lambda_{2}-\lambda_{6})) \, \nonumber \\  \label{NCM}  \\ \nonumber
 M^{2}_{h^{+}} & = & \frac{\lambda_{8}+\frac{1}{2} }{2}(v^{2}+v_{\chi}^{2})
\eea
where the constants $\lambda_{2}$, $\lambda_{6}$ and $\lambda_{8}$ are coupling constants of the scalar potential (see Eq.6 of \cite{Mizukoshi:2010ky}.). In order to obtain analytical expressions for the masses of the scalars simplifying assumptions were made concerning some couplings. We emphasize that they do not affect our conclusions. 

Summarizing, the contributions of the {\it 3-3-1 r.h.n} model to $a_\mu$ rise from the neutral (FIG. \ref{feynman7}) and singly-charged gauge bosons (FIG. \ref{feynman9}), the neutral (FIG. \ref{feynman1}) and charged scalars (FIG. \ref{feynman3}). The main difference between this model and the {\it Minimal 3-3-1} model is the absence of a doubly charged gauge boson which was the main player in the {\it Minimal 3-3-1} model.  

\section{3-3-1 LHN}

The 3-3-1 model with heavy neutral lepton (3-3-1 LNH) \cite{Mizukoshi:2010ky,Catano:2012kw} differs from the  {\it 3-3-1 r.h.n} because now the right-handed neutrino is replaced by a heavy neutral lepton ($N$) whose mass is governed by the scale of symmetry breaking of the 3-3-1 symmetry. The lepton generations are arranged as follows,

\begin{equation} f^{a}_L =
\begin{pmatrix}
   \nu^a  \\
    l^a  \\
    N^a
\end{pmatrix} ;\,\, l^a_R, N^a_R.
\label{lrr1}
\end{equation}

The same scalar triplets defined in Eq.\ref{tripletscalars2} appear in this model. Consequently, the $Z'$ and $W'$ masses are precisely the same as Eqs.\eqref{zprimemassmin} and (\ref{wprimemassmin}) respectively. The neutral current is also the same. There is a subtle but important difference that resides in the charged current, 
\begin{eqnarray}
{\cal L} \supset - \frac{g}{\sqrt{2}}\left[
\overline{N_L}\, \gamma^\mu \bar{l_L}\, W^{\prime -}_\mu  \right].
\label{heavylep1}
\end{eqnarray}

The vector and vector-axial couplings in the charged current that appears in Eq.\ref{heavylep1} are easily extracted. Similarly to the {\it 3-3-1 r.h.n} this model will have interactions involving scalars of the form,

\begin{eqnarray}
{\cal L} \supset && G_l\, \overline{l_R}\, N_L h_1^- + G_l\, \overline{l_R}\, \nu_L h_2^+ + G_s \bar{\mu}\, \mu S_2,
\label{Llhn}
\end{eqnarray}where again $G_s=m_{\mu} \sqrt{2}/(2 v)$, and $G_l=m_l\sqrt{2}/v_\eta$. We point out that the scalar fields $h_1^-$ and $h_2^-$ are the mass eingenstates that arise of the diagonalization procedure of the $(\chi^-,\rho'^-)$ and $(\eta^-,\rho^-)$ bases  and $S_2$ is a combination of $R_\eta$ and $R_\rho$\cite{Mizukoshi:2010ky}. Their masses are found to be  \cite{Mizukoshi:2010ky,Catano:2012kw},
\begin{eqnarray}
M^{2}_{h^{-}_{1}} & = & \frac{\lambda_{8}+\frac{1}{2} }{2}(v^{2}+v_{\chi}^{2})\,, \nonumber \\
M^{2}_{h^{-}_{2}} & = & \frac{v_{\chi}^{2}}{2}+\lambda_{9}v^{2}\,,\nonumber\\
M^{2}_{S_{2}} & = & \frac{1}{2}(v_{\chi}^{2}+2v^{2}(2\lambda_{2}-\lambda_{6})),
\label{massash1h2}
\end{eqnarray} where $\lambda_{2}$, $\lambda_{8}$, $\lambda_{9}$ are the coupling constants of the scalar potential in this model (See Eq.(6) of \cite{Mizukoshi:2010ky}). As always, $v_\chi$ is assumed to be much larger than $v_\rho$ and $v_\eta$, as $v_\rho^2+v_\eta^2 = 246^2$ GeV$^2$. For simplicity we take $v_\rho=v_\eta = v/\sqrt{2}$. We highlight that this assumption does not interfere in our conclusions as the scalar contributions are suppressed anyway in comparison with the ones stemming from gauge bosons.

In summary, in the  3-3-1 LHN we have a contributing coming from the $S_2$  ( FIG. \ref{feynman1} ), $h_2^-$ ( FIG. \ref{feynman3} ), $h_1^-$ ( FIG. \ref{feynman4} ), $Z^\prime$  (FIG. \ref{feynman7} ) and $W^\prime$ ( FIG. \ref{feynman10} ).

\section{Economical 3-3-1 Model}

The  {\it Economical 3-3-1} model \cite{Dong:2006mg,Dong:2008ya,Berenstein:2008xg,Martinez:2014lta} is built with the same leptonic triplet of the {\it  3-3-1 r.n.h} model, but now with scalar sector reduced to incorporate two scalar triplets, only:

\begin{equation}
    \chi=\begin{pmatrix}
         \chi_1^0\\
         \chi_2^{-}\\
         \chi_3^0
    \end{pmatrix},
    \eta=\begin{pmatrix}
         \eta_1^+\\
         \eta_2^0\\
         \eta_3^+
    \end{pmatrix},
    \label{tripletscalars3}
\end{equation}where the $vev$ of neutral fields takes the form $\left<\eta^0_2\right>=v_{\eta^0_2}=v/\sqrt{2}$, $\left<\chi^0_1\right>= v_{\chi_1}=u/\sqrt{2}$, $\left<\chi^0_3\right>=v_{\chi^0_3}= v_{\chi}/\sqrt{2}$. Once again we assume that $u,v \ll v_{\chi}$.

The corrections for $a_\mu$ arise from neutral and charged scalars via Yukawa Lagrangian \cite{Dong:2006mg},
\begin{equation}
{\cal L} \supset G^\ell_{ij}\overline{f}_{iL}\eta \ell_{jR}+ G^\epsilon_{ij}\epsilon_{pmn}(\overline{f}^c_{iL})_p(f_{iL})_m (\eta)_n 
+ h.c.
\label{lyukawa}
\end{equation}

From this Lagrangian we  will obtain terms that go with $G_l \bar{l_R} \nu_L \eta_1^+$ and $G_s \bar{\mu} \mu S_2$, where,
\bea
M_{\eta_1^+}^2 =\frac{ \lambda_4}{2} \left( u^2 +v^2 + v_{\chi}^2 \right), \
M_{S_2}^2 = 2 \lambda_1 v_{\chi}^2.
\label{meta}
\eea

The neutral and charged currents are the same as the  {\it 3-3-1 r.h.n}, but the masses of the gauge bosons  take different forms because we now have only two scalar triplets,

\begin{equation}
M_{Z^{\prime}}^2 \approx \frac{g^2 c^2_W v_{\chi}^2}{3-4s_W^2}, \ \, \ M_{W^{\prime}}^2 = \frac{g^2}{4}\left( v_{\eta}^2 + v_{\chi}^2 \right).
\label{mzprime}
\end{equation}

The {\it Economical 3-3-1} features basically the same new physics contributions to $a_\mu$ as the {\it 3-3-1 r.h.n}. The main differences appear in the expressions for the $W^\prime$ and $Z^\prime$ masses. As we are going to plot the overall contribution of each 3-3-1 model to $a_\mu$ as a function of the $v_{\chi}$, as we will see, the results will differ.


\section{3-3-1 Model with exotic Leptons}

3-3-1 models are known for being versatile, with many possible combinations of fermionic and scalar multiplets that can be constructed in an anomaly-free way. However, the price that usually has to be paid is that the quarks generations cannot be represented equally under $SU(3)_L$. This happens in the {\it 3-3-1 model with exotic leptons}. A consequence of this necessity is that $Z^{\prime}$ interactions are no longer universal, leading to flavor changing neutral currents at tree level \cite{Cogollo:2013mga}. In this section, the goal is to investigate models that have generations with different representations under $SU(3)_L$ in the leptonic sector. Here we are going to consider the following leptonic sector \cite{Ponce:2001jn}

\begin{equation} f_{1L} =
\begin{pmatrix}
   \nu_1  \\
    l_1  \\
    E_1^-
\end{pmatrix} ;\,\, l_1^c;  f_{2,3L} =
\begin{pmatrix}
   \nu_{2,3}  \\
    l_{2,3}  \\
    N_{2,3}
\end{pmatrix} ;\,\, l_{2,3}^c; 
\end{equation}
\begin{equation} 
f_{4L}=
\begin{pmatrix}
   E_2^-  \\
    N_3  \\
    N_4
\end{pmatrix} ;\,\, E_{2}^c;\\
f_{5L}=
\begin{pmatrix}
   N_5 \\
    E_3^+  \\
    l_3^+
\end{pmatrix} ;\,\, E_{3}^c;
\label{lrr2}
\end{equation}where $N$ and $E$ are the exotic neutral and charged leptons, respectively.


We label the new gauge bosons as $K^{\pm}$, $K^0$ and $Z^{\prime}$. The scalar sector is formed by three triplets, given by \cite{Ponce:2001jn,Ponce:2002fv,Anderson:2005ab,Cabarcas:2013jba},

\begin{equation}
    \chi_i=\begin{pmatrix}
         \chi_i^-\\
         \chi_i^{0}\\
         \chi_i^{0\prime}
    \end{pmatrix},
    \chi_3=\begin{pmatrix}
         \chi_3^0\\
         \chi_3^+\\
         \chi_3^{\prime+}
    \end{pmatrix}.
    \label{tripletscalars4}
\end{equation} with $i=1,2$, $\left<\chi_1\right> = \left(0,\  0,\ v_{\chi}\right)^T$, $\left<\chi_2\right> = \left(0,\  v/\sqrt{2},\ 0\right)^T$ and $\left<\phi_3\right> = \left(v^{\prime}/\sqrt{2},\  0,\ 0\right)^T$, where $v_{\chi} \gg v,v^\prime$, with $v^{\prime} \sim v$.

The relevant interactions to $a_\mu$ are \cite{Cabarcas:2013jba},
\begin{eqnarray}
{\cal L} \supset &&\frac{g^{\prime}}{2 \sqrt{3} s_W c_W} \bar{\mu}\gamma_{\mu} \left( g_V + g_A \right) \mu \, Z^{\prime}\nonumber\\
&& - \frac{g}{\sqrt{2}} \left( \overline{N_{1L}}\, \gamma_{\mu} \mu_{L} + \bar{\mu}_L
 \gamma_{\mu} N_{4L} \right) K^{+}_{\mu}\nonumber\\
 &&-\frac{g}{\sqrt{2}} \left( \bar{\mu}_L \gamma_{\mu} E_L \right) K^0_{\mu}+h_1 \bar{\mu} (1-\gamma_5) N \chi^+\nonumber\\
&& + h_2 \bar{\mu} E^- \chi^0 +h_3 \bar{\mu} E_2^- \chi^0  + \mbox{H.c.},
\label{exotcontri}
\end{eqnarray}
where $\chi^+$ and $\chi^0$ are scalars coming from the scalar triplets. The vector and vector-axial couplings of the $Z^{\prime}$ to charged leptons are,
\begin{equation}
g_V = \frac{-c_{2W} + 2 s_W^2}{2}, \, g_A=\frac{c_{2W} + 2 s_W^2}{2},
\label{gvga}
\end{equation}  and the masses of the gauge bosons are found to be \cite{Cabarcas:2013jba},
\bea
M_{Z^{\prime}}^2 & = &\frac{2}{9} \left( 3 g^2 + g^{\prime 2} \right) v_{\chi}^2,\nonumber\\
M_{K^+}^2 & =M_{K^0}^2= & \frac{g^2}{4} \left( 2 v_{\chi}^2 + v^2 \right),\nonumber\\
g^{\prime} & = &\frac{g \tan_W}{\sqrt{1- \tan_W^2/3}}.
\label{mzwexoc}
\eea 

As highlighted previously, the corrections to $a_\mu$ coming from the scalars are suppressed by the lepton masses, so they will not be considered in this case. Hence, the main corrections to $a_\mu$ rise from the gauge bosons $Z^{\prime}$ ( FIG. \ref{feynman7}), $K^0$  ( FIG. \ref{feynman8}) and $K^{-}$  ( FIG. \ref{feynman11}). The Feynman diagrams induced by the presence of $K^{-}$ and $K^0$ also involve an exotic charged lepton $E$, whose mass can be very large. 


\section{Contributions for the Muon Anomalous Magnetic Moment}

Having in mind the relevant interactions for \gmu\, appearing in 3-3-1 models, in this section we provide general contributions to \gmu\, which will be used later taking into account the particularities of each 3-3-1 model.
\subsubsection{Neutral Scalar Mediator}

A new scalar ($\phi$) may induce two possible corrections to \gmu\, as represented in Figs.~\ref{feynman1} and \ref{feynman2}. In this case, the general expression for $\Delta a_\mu$ is written as follows,

\begin{widetext}

\begin{eqnarray}
\Delta a_\mu(\phi) = \frac{1}{8\pi^2}\frac{m_\mu^2}{M_\phi^2} \int_0^1 \mathrm{d}x\,\sum_f \left[\frac{\left|g_{s\,1}^{f\mu}\right|^2 P_1^+(x)+\left|g_{p\,1}^{f\mu}\right|^2 P_1^-(x)}{(1-x)(1-x \lambda^2)+x\,\epsilon_f^2\lambda^2} \right],
 \label{gmu_NS}
\end{eqnarray}
  
where
\begin{equation}
  P_1^\pm(x) = x^2\left(1-x \pm \epsilon_f\right), \label{P1}
\end{equation}
with $g_{s\,1}^{f\mu}$ and $g_{p\,1}^{f\mu}$ being the scalar (s) and pseudo-scalar (p) matrices in flavor space, $\epsilon_f\equiv \frac{m_f}{m_\mu}$ and $\lambda\equiv \frac{m_\mu}{M_\phi}$. The fermion mass is $m_f=m_\mu$ for the Fig. \ref{feynman1} and $m_f=m_E$ for the Fig. \ref{feynman2}, with $M_\phi$ being the scalar mass.
It is important to mention that in the limit of heavy mediator, $M_\phi \gg m_\mu,m_f$, the analytical expression for $\Delta a_\mu$ is simplified to,

\begin{eqnarray} \label{gmu_NS_approx}
\Delta a_\mu(\phi) \simeq \frac{1}{4\pi^2}\frac{m_\mu^2}{M_\phi^2} \sum_f \left[\left|g_{s\,1}^{f\mu}\right|^2 \left( \frac{1}{6} - \epsilon_f\left(\frac{3}{4} + \log(\epsilon_f\lambda)\right)\right)+ \left|g_{p\,1}^{f\mu}\right|^2 \left( \frac{1}{6} + \epsilon_f\left(\frac{3}{4} + \log(\epsilon_f\lambda)\right) \right)  \right].
\end{eqnarray}

\subsubsection{Singly Charged Scalar Mediator}
A singly charged scalar ($\phi^\pm$) generates corrections to the \gmu\, via Figs.~\ref{feynman3}-\ref{feynman4}, where $\nu_{L}$ are the SM neutrinos, and $N$ heavy neutral leptons.  If lepton number is not conserved a charged operator might be present in the Lagrangian. Either way, the general result for $\Delta a_\mu$ is found to be, 

  \begin{equation}
    \Delta a_\mu\left(\phi^+\right) = \frac{-1}{8\pi^2} \frac{m_\mu^2}{M_{\phi^+}^2}\int_0^1 \mathrm{d}x\,\sum_f \frac{\left|g_{s\,2}^{f\mu}\right|^2 P_2^+(x)+\left|g_{p\,2}^{f\mu}\right|^2 P_2^-(x)}{\epsilon_f^2\lambda^2(1-x)\left(1-\epsilon_f^{-2}x\right)+x},\label{gmu_CS}
  \end{equation}

  where
  \begin{equation}
    P_2^\pm(x) = x(1-x)\left(x \pm \epsilon_f\right),\quad
    \label{P2} 
  \end{equation}
with $\epsilon_f\equiv \frac{m_{\nu}}{m_\mu}$ and $ \lambda\equiv \frac{m_\mu}{M_{\phi^+}}$. In the heavy scalar mediator regime we obtain, 
\begin{eqnarray}\label{gmu_SS_approx}
	\Delta a_\mu(\phi^+) \simeq  \frac{-1}{4\pi^2}\frac{m_\mu^2}{M_{\phi^+}^2} \sum_f \left[ \left|g_{s\,2}^{f\mu}\right|^2 \left( \frac{1}{12} + \frac{\epsilon_f}{4} \right) + \left|g_{p\,2}^{f\mu}\right|^2 \left( \frac{1}{12} - \frac{\epsilon_f}{4} \right) \right].
\end{eqnarray}

\subsubsection{Neutral Gauge Boson Mediator}

The neutral gauge boson corrections to the \gmu\, are given by the diagrams of Figs. \ref{feynman7}-\ref{feynman8}. In some cases we have an exotic charged fermion, $E$, and when that happens we label the neutral gauge boson $K^0$ in addition to the $Z^\prime$. The general contribution for $\Delta a_\mu$ is given by, 

 \begin{eqnarray}
     \Delta a_\mu\left(E, Z^\prime\right) &=& \frac{1}{8\pi^2} \frac{m_\mu^2}{M_{Z^\prime}^2} \int_0^1 \mathrm{d}x \sum_f \left[ \frac{\left|g_{v\,1}^{f\mu}\right|^2 P_3^+(x) +\left|g_{a\,1}^{f\mu}\right|^2 P_3^-(x)}{(1-x)\left(1-\lambda^2 x\right)+\epsilon_f^2\lambda^2 x} \right],
    \label{gmu_NGBM} 
 \end{eqnarray}

  with 
  \begin{equation}
    P_3^\pm = 2x(1-x) (x-2\pm 2\epsilon_f) + \lambda^2x^2(1 \mp \epsilon_f)^2(1-x \pm \epsilon_f),
    \label{P4}
  \end{equation}where $\epsilon_f \equiv  \frac{m_{E_f}}{m_\mu}$, $\lambda \equiv \frac{m_\mu}{M_{Z^\prime}}$, with $g_{v\,1}^{f\mu}$ and $g_{a\,1}^{f\mu}$ being the vector and vector-axial coupling constants. Considering the neutral boson much heavier than the fermions we obtain a simplified expression, 
  
\begin{eqnarray} \label{neutralGBapprox}
	\Delta a_\mu\left(E,Z^\prime\right) \simeq \frac{-1}{4\pi^2}\frac{m_\mu^2}{M_{Z^\prime}^2} \sum_f \left[ \left|g_{v\,1}^{f\mu}\right|^2 \left( \frac{2}{3} - \epsilon_f \right) + \left|g_{a\,1}^{f\mu}\right|^2 \left( \frac{2}{3} + \epsilon_f \right) \right].
\end{eqnarray}

Note that in $\epsilon_f \equiv 1$ in the absence of exotic fermions. In this case, vector neutral gauge bosons give rise to a positive contribution to g-2. Generally speaking, the overall sign depends on the relative strength between the vector and vector-axial couplings.

\subsubsection{Charged Gauge Boson Mediator}

The Feynman diagrams in Figs. \ref{feynman9}- \ref{feynman10}-\ref{feynman11} account for the possible contributions stemming from a singly charged gauge boson ($W^\prime$ or $K$) and a neutral fermion ($\nu$ or $N$). In this case, the more general expression for $\Delta a_\mu$ is given by,

\begin{eqnarray}
  \label{gmu_CGB}
    \Delta a_\mu\left(N,W^\prime\right) &=& \frac{-1}{8\pi^2} \frac{m_\mu^2}{M_{W^\prime}^2} \int_0^1 \mathrm{d}x \sum_f  \frac{\left|g_{v\,2}^{f\mu}\right|^2 P_4^+(x)+ \left|g_{a\,2}^{f\mu}\right|^2 P_4^-(x)}{\epsilon_f^2\lambda^2(1-x)\left(1-\epsilon_f^{-2}x\right)+x},
\end{eqnarray}

  with
 \begin{equation}\label{P3}
    P_4^\pm = -2 x^2 (1 + x \mp 2\epsilon_f) + \lambda^2 x (1-x) (1 \mp \epsilon_f)^2\left(x \pm \epsilon_f\right)
  \end{equation}where 
and $\epsilon_f \equiv \frac{m_{N_f}}{m_\mu}$,  $\lambda \equiv \frac{m_\mu}{M_{W^\prime}}$. $g_v$ and $g_a$ are again the vector and vector-axial couplings. In 3-3-1 models, we have either $m_{N_f} = m_{\nu}$ or $m_{N_f}= m_{N}$ following Figs. \ref{feynman9}-\ref{feynman10}-\ref{feynman11}. $M_{W^\prime}$ is the charged gauge boson mass. Taking $M_{W^\prime}\gg m_N,m_f$ the overall correction to \gmu simplifies to,
  
  \begin{eqnarray}
      \Delta a_\mu(N,W^\prime) \simeq \frac{1}{4\pi^2}\frac{m_\mu^2}{M_{W^\prime}^2} \sum_f \left[ \left|g_{v\,2}^{f\mu}\right|^2 \left( \frac{5}{6} - \epsilon_f\right) + \left|g_{a\,2}^{f\mu}\right|^2 \left( \frac{5}{6} + \epsilon_f\right) \right].
  \label{approxamu_W1}
  \end{eqnarray}

\subsubsection{Doubly Charged Vector Boson Mediator}

Doubly charged vector bosons ($U^{\pm\pm}$) are rare fields in model building endevours but common in some 3-3-1 model constructions. The Feynman diagrams that contribute to the \gmu\, are displayed in Figs.\ref{feynman12}-\ref{feynman13} and their contribution are given by,

\begin{align}
      \Delta a_\mu\left(U^{++}\right) = \frac{8}{8\pi^2} \frac{m_\mu^2}{M_U^2} \int_0^1 \mathrm{d}x \sum_f \frac{\left|g_{v\,3}^{f\mu}\right|^2 P_4^+(x) + \left|g_{a\,3}^{f\mu}\right|^2 P_4^-(x)}{\epsilon_f^2\lambda^2(1-x)\left(1-\epsilon_f^{-2}x\right)+x} -\frac{4}{8\pi^2} \frac{m_\mu^2}{M_U^2} \int_0^1 \mathrm{d}x \sum_f \frac{\left|g_{v\,3}^{f\mu}\right|^2 P_3^+(x) + \left|g_{a\,3}^{f\mu}\right|^2 P_3^-(x)}{(1-x)\left(1-\lambda^2 x\right)+\epsilon_f^2\lambda^2 x},
  \label{gmu_DV} 
\end{align}where $\epsilon_f\equiv\frac{m_f}{m_\mu}$, $\lambda\equiv\frac{m_\mu}{M_U}$, and $g_{a\,3}^{f\mu}$ ($g_{v\,3}^{f\mu}$) are symmetric and anti-symmetric couplings in flavor space. In the case that $f=\mu$, we will have identical fields, thus $g_{v\,3}^{\mu\mu}=0$ \cite{Lindner:2016bgg}. In the limit $M_{U} \gg m_f,m_{\mu}$) we get,

\begin{eqnarray}
  \Delta a_\mu\left(U^{++}\right) \simeq \frac{1}{\pi^2} \frac{m_\mu^2}{M_U^2} \sum_f \left[ \left|g_{v\,3}^{f\mu}\right|^2 \left[ -1 + \epsilon_f\right] - \left|g_{a\,3}^{f\mu}\right|^2 \left[ 1 + \epsilon_f\right]\right].
  \label{gmu_DV_limit} 
\end{eqnarray}

We highlight that for $\epsilon_f=1$ the overall doubly charged gauge boson contribution is negative and the vector current is null.

\subsection{Minimal 3-3-1 Model}

The relevant interactions for the minimal 3-3-1 model were discussed section \ref{minimal}. The corrections to \gmu rise from the presence of new gauge bosons $U^{++}$, $Z^\prime$ and $W^\prime$, and charged scalar $\eta_1^-$.  The parameters that were used in this model will be explained below, considering the general analytical expressions given above.
\begin{itemize}
    \item The case $U^{++}$ contribution is based on \e \ref{gmu_DV}, where $g_{a\,3}^{\mu\mu}=-\frac{g}{2\sqrt{2}}$ and $g_{v\,3}^{\mu\mu}=0$.  So, we should get $\Delta a_{\mu}\left(U^{++}\right)$, where a muon and a doubly charged boson are the mediators for the \gmu process. In the limit $M_{U} \gg m_{\mu}$ we obtain,

\begin{eqnarray}
  \Delta a_\mu\left(U^{++}\right) \simeq -2\frac{1}{\pi^2} \frac{m_\mu^2}{M_U^2}  \left|\frac{g}{2\sqrt{2}}\right|^2 .
  \label{gmu_DV_limit1} 
\end{eqnarray}

Notice that the correction to \gmu\, is indeed negative and simply governed by $M_U$.

    \item The $W^\prime$ correction to \gmu\, is obtained from \e ~\ref{gmu_CGB}. We should take $\epsilon_f \equiv \frac{m_\nu}{m_\mu}$, $g_{v\,2}^{\nu\mu} = g_{a\,2}^{\nu\mu} = \frac{g}{2\sqrt{2}}$. In the heavy mediator regime we get,
  
  \begin{eqnarray}
      \Delta a_\mu(\nu,W^\prime) \simeq \frac{1}{4\pi^2}\frac{m_\mu^2}{M_{W^\prime}^2} \left|\frac{g}{2\sqrt{2}}\right|^2   \left( \frac{5}{3}  \right).
      \label{Weqlimit}
  \end{eqnarray}
  
We point out that the result in Eq.\ref{Weqlimit} cannot be used when we have a neutral fermion N present with $m_N$ being sufficiently large. In that case one needs to numerically solve Eq.\ref{gmu_CGB}.
    
    \item The $Z^\prime$ correction to \gmu\, is obtained from \e \ref{gmu_NGBM}. Using $m_{E_f}=m_\mu$, taking vector and vector-axial couplings from \e \ref{hsm} we find the heavy $Z^\prime$ limit, 
    
\begin{eqnarray} 
	\Delta a_\mu\left(\mu,Z^\prime\right) \simeq \frac{-1}{4\pi^2}\frac{m_\mu^2}{M_{Z^\prime}^2} \left|\frac{g}{2c_W}
\frac{\sqrt{3}\sqrt{1 - 4 s_W^2}}{2}\right|^2\left( -\frac{4}{27}\right).
\end{eqnarray}
    
    \item Concerning the charged scalar contribution, we can derive it using \e \ref{gmu_CS}, with $\epsilon_f\equiv \frac{m_{\nu}}{m_\mu}$ and $g_{s2}^{\nu\mu}=g_{p2}^{\nu\mu}=\frac{m_\mu\sqrt{2}}{2 v_\eta}$. For $M_{\eta_1^+} \gg m_\mu,m_{\nu_L}$ we obtain,
\begin{eqnarray}
	\Delta a_\mu(\eta_1^+) \simeq  \frac{-1}{4\pi^2}\frac{m_\mu^2}{M_{\eta_1^+}^2}  \left|\frac{m_\mu\sqrt{2}}{2 v_\eta}\right|^2 \left( \frac{1}{6}\right) .
\end{eqnarray}
\end{itemize}

Notice that the correction to \gmu\, is negative and suppressed the muon mass. This occurs because the charged scalar is embedded in a scalar multiplet that gets a non-trivial vacuum expectation value. This means that the scalars in the multiplet interaction with fermions proportionally to their masses.

\subsection{3-3-1 r.h.n}

We will use the same logic of the 3-3-1 minimal model to find the individual corrections to \gmu. To avoid being repetitive we simply quote them in the heavy mediator regime. With that being said, we found,
  
  \begin{eqnarray}
      \Delta a_\mu(\nu,W^\prime) \simeq \frac{1}{4\pi^2}\frac{m_\mu^2}{M_{W^\prime}^2} \left|\frac{g}{2\sqrt{2}}\right|^2   \left( \frac{5}{3}  \right),
  \end{eqnarray}
  
\begin{eqnarray} 
	\Delta a_\mu\left(\mu,Z^\prime\right) \simeq \frac{-1}{4\pi^2}\frac{m_\mu^2}{M_{Z^\prime}^2} \frac{1}{3}\left|-\frac{g}{4 c_W \sqrt{3-4s_W^2}}\right|^2   \left[ -\left|  1 -
4 s_W^2\right|^2  +5  \right],
\end{eqnarray}
    
\begin{eqnarray}
	\Delta a_\mu(h^+) \simeq  \frac{-1}{4\pi^2}\frac{m_\mu^2}{M_{h^+}^2}  \left|\frac{m_\mu\sqrt{2}}{2 v_\eta}\right|^2  \frac{1}{6},
\end{eqnarray}
    
\begin{eqnarray} 
\Delta a_\mu(S_2) \simeq \frac{1}{4\pi^2}\frac{m_\mu^2}{M_{S_2}^2}  \left(\frac{m_\mu\sqrt{2}}{2 v_\eta}\right)^2 \left[ \frac{1}{6} - \left(\frac{3}{4} + \log\left(\frac{m_\mu}{M_{S_2}}\right)\right)\right].
\end{eqnarray}

In summary, the 3-3-1 r.h.n yields four contributions to g-2.

\subsection{3-3-1 LHN}

This models induces several corrections to \gmu, coming from the $Z^\prime$, $W^\prime$, $h_1^-$ and $h_2^-$ and $S_2$. The contributions to \gmu\, rising from the $Z^\prime$, $h_1^-$ and $S_2$ fields are identical to the the 3-3-1 r.h.n model. The $W^\prime$ contribution is a bit different because of the presence of neutral heavy fields, $N$. In the previous model, this field was replaced by the right-handed neutrino. Hence, we only exhibit the different corrections to \gmu. Taking $\epsilon_f \equiv \frac{M_N}{m_\mu}$, $g_{v\,2}^{N\mu} = g_{a\,2}^{N\mu} = \frac{g}{2\sqrt{2}}$ we get,
    
  \begin{eqnarray}
      \Delta a_\mu(N,W^\prime) \simeq \frac{1}{4\pi^2}\frac{m_\mu^2}{M_{W^\prime}^2} \left|\frac{g}{2\sqrt{2}}\right|^2   \frac{5}{3}
  \end{eqnarray}and for the charged scalar $h_2$,

\begin{eqnarray}
	\Delta a_\mu(h_2^+) \simeq  \frac{-1}{4\pi^2}\frac{m_\mu^2}{M_{h_2^+}^2} \left|\frac{m_\mu\sqrt{2}}{2 v_\eta}\right|^2 \frac{1}{6} .
	\label{eqh2}
\end{eqnarray}

The scalar and pseudo-scalar coupling of this charged scalar are equal to $m_\mu \sqrt{2}/(2 v_\eta)$ and for this reason we observe a an extra muon mass dependence on Eq.\ref{eqh2}.

\subsection{Economical 3-3-1}

In this model, the new contributions to \gmu\, are obtained by studying the processes mediated by $Z^\prime$, $W^\prime$, $\eta_1^+$ and $S_2$. The corrections to $\Delta a_\mu$ that arise from  $Z^\prime$ and $W^\prime$ have nearly the same magnitude as in the 3-3-1 r.h.n model. The main difference is in the gauge boson masses that have a slightly different dependence with the energy scale at which the 3-3-1 symmetry is broken. It is straightforward to find that, 

\begin{eqnarray}
	\Delta a_\mu(\eta_1^+) \simeq  \frac{-1}{4\pi^2}\frac{m_\mu^2}{M_{\eta_1^+}^2} \left|\frac{m_\mu\sqrt{2}}{2 v_\eta}\right|^2 \frac{1}{6} ,
\end{eqnarray}

\begin{eqnarray} 
\Delta a_\mu(S_2) \simeq \frac{1}{4\pi^2}\frac{m_\mu^2}{M_{S_2}^2}  \left(\frac{m_\mu\sqrt{2}}{2 v_\eta}\right)^2 \left[ \frac{1}{6} - \left(\frac{3}{4} + \log\left(\frac{m_\mu}{M_{S_2}}\right)\right)\right].
\end{eqnarray}

\subsection{3-3-1 with exotic leptons}

The corrections to \gmu\, stemming from the 3-3-1 with exotic leptons model are governed by the $Z^\prime$, $K^0$ and $K^+$ bosons. In summary we get,
  \begin{eqnarray}
      \Delta a_\mu(N,K^+) \simeq \frac{1}{4\pi^2}\frac{m_\mu^2}{M_{K^+}^2} \left|\frac{g}{\sqrt{2}}\right|^2   \frac{5}{3},
  \end{eqnarray}
   
\begin{eqnarray} 
	\Delta a_\mu\left(E,K^0\right) \simeq \frac{-1}{4\pi^2}\frac{m_\mu^2}{M_{K^0}^2}  \left|\frac{g}{\sqrt{2}}\right|^2 \left( \frac{4}{3} \right),
\end{eqnarray}

\begin{eqnarray} 
	\Delta a_\mu\left(\mu,Z^\prime\right) \simeq \frac{-1}{4\pi^2}\frac{m_\mu^2}{M_{Z^\prime}^2} \left|\frac{g^{\prime}}{2 \sqrt{3} s_W c_W}\right|^2 \frac{1}{12} \left[ -\left| \left(-c_{2W} + 2 s_W^2\right)\right|^2 + 5 \left|\left(c_{2W} + 2 s_W^2\right)\right|^2  \right].
\end{eqnarray}

To sum up, the possible analytical expressions for the corrections to \gmu ~from models 3-3-1 were described in this section. Where their numerical results will be shown in the section \ref{results}.

\end{widetext} 
\section{Results}
\label{results}
In this section, we will present the main results of our work and put them into perspective with collider bounds. In all models discussed here, the leading corrections to $a_\mu$ stem from the new gauge bosons. As we are dealing with a $SU(3)_L$ gauge group there are several gauge bosons that play a role in $a_\mu$. We have computed all individual corrections to $a_\mu$ and expressed our results in terms of the scale of symmetry breaking, $v_{\chi}$. For completeness we display the individual contributions in the Appendix. As the gauge bosons and scalar fields have masses governed by $v_{\chi}$ we can sum up all the individual contributions and draw conclusions in terms of the total $\Delta a_\mu$ for each model. The main results are summarized in the {\it Table} \ref{bosonmassbounds}. We start discussing our findings in the context of the {\it Minimal 3-3-1} model.

\subsection{Minimal 3-3-1} 
\begin{figure*}[!ht]
\centering
\includegraphics[scale=0.8]{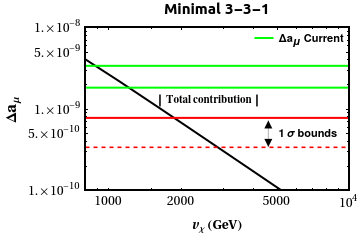}
\caption{Overall contribution to $\Delta a_\mu$ from the {\it Minimal 3-3-1} model. The green bands are delimited by $\Delta a_\mu=(261\pm 78)\times 10^{-11}$. The current $1\sigma$ bound is found by requiring $\Delta a_\mu< 78 \times 10^{-11}$ while the projected bound is obtained for $\Delta a_\mu< 34 \times 10^{-11}$. We used $M_{Z^\prime}=0.395 v_{\chi}$, $M_{W^\prime}=M_{U^{\pm\pm}}=0.33 v_{\chi}$.  } \label{Graph1}
\end{figure*}

In FIG. \ref{Graph1} we show the total contribution to $\Delta a_\mu$ as a function of $v_\chi$ for the {\it Minimal 3-3-1} model with a black curve. There we also display a green band at which the $a_\mu$ anomaly is addressed. Moreover, we exhibit the current and projected $1\sigma$ bounds considering the precision aimed by the g-2 experiment at FERMILAB.  It is important to mention that the individual contributions were calculated numerically considering $v_\eta=v_\rho =147$~GeV. The $Z^\prime$ correction to $a_\mu$ is small. The largest corrections to $a_\mu$ stem from the singly ($W'$) and doubly ($U^{++}$) charged gauge bosons. The latter gives rise to a negative contribution to $a_\mu$. This is an important point that has been overlooked in \cite{Ky:2000ku,Kelso:2014qka}. The vector current vanishes for a doubly charged gauge boson  \cite{Lindner:2016bgg}.  The main contributions come from the $W'$ and $U^{++}$ gauge bosons in the {\it Minimal 3-3-1} model, which are positive and negative respectively. The overall correction to $\Delta a_\mu$ is small and negative. Looking at FIG. \ref{Graph1} one can easily conclude that the $v_{\chi} > 1.8$~TeV, and using the projected g-2 sensitivity we will be able to impose $v_\chi > 2.8$~TeV. This model cannot explain $\Delta a_\mu$ as it yields a negative correction to $a_\mu$. Moreover, one should keep in mind the existing collider bounds which were placed on the mass of the new gauge bosons (See {\it Table} \ref{bosonmassbounds}). One may use the relations $M_{Z^\prime}=0.395 v_{\chi}$, and $M_{W^\prime}=M_{U^{\pm\pm}}=0.33 v_{\chi}$  to obtain constraints on the scale of symmetry breaking. Currently LHC data imposes $M_{Z^\prime} > 3.7$~TeV which means $v_{\chi} > 9.3$~TeV. Having in mind the Landau pole found at 5TeV, the {\it Minimal 3-3-1} model has been excluded by LHC searches. There are ways to extend the {\it Minimal 3-3-1} model and add extra decay channels to the $Z^\prime$ gauge boson \cite{Machado:2016jzb}, without altering the g-2 predictions. Such new decay modes may weaken the LHC bound. Therefore it is worthwhile to determine the contributions to $a_\mu$ as it gives rise to an orthogonal and complementary bound on the model. 

We point out that each contribution to $a_\mu$ depends on the mass of the particles running in the loop. However, we can describe each correction to $a_\mu$ in terms of the scale of symmetry breaking, $v_{\chi}$, as the particle masses depend on it. In this way, we obtain the lower mass bounds on the masses of the gauge bosons using FIG. \ref{Graph1}. We summarize these bounds in {\it Table} \ref{bosonmassbounds}. 

In summary, if the $a_\mu$ anomaly is confirmed by the g-2 experiment, the {\it Minimal 3-3-1} model cannot offer an answer. New interactions with charged leptons must arise.

\begin{figure*}[!ht]
\centering
\includegraphics[scale=0.8]{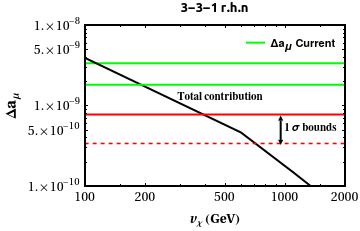}
\caption{Overall contribution to $\Delta a_\mu$ from the {\it 3-3-1 r.h.n} model. We used $M_{Z^\prime}=0.395 v_{\chi}$, $M_{W^\prime}=M_{U^{\pm\pm}}=0.33 v_{\chi}$. } \label{Graph2}
\end{figure*}

\begin{figure*}[!ht]
\centering
\includegraphics[scale=0.7]{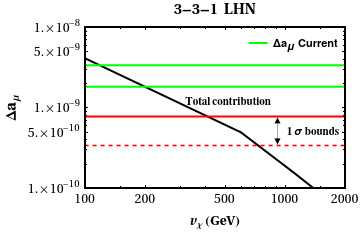}
\includegraphics[scale=0.7]{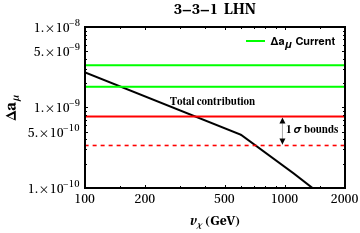}
\caption{Overall contribution to $\Delta a_\mu$ from the {\it 3-3-1 LHN} model for $m_N = 1$~GeV (left-panel) and $m_N = 100$~GeV (right-panel). One can clearly from the plots that our conclusions concerning the {\it 3-3-1 LHN} heavily depend on mass used for the neutral lepton. For $m_N=100$~GeV we can place a projected limit of $v_\chi > 6.4$~TeV, whereas for $m_N \leq 240$~GeV no limit on the scale of symmetry breaking can be found because the corrections to $\Delta a_\mu$ is too small. We used $M_{Z^\prime}=0.395 v_{\chi}$,$M_{W^\prime}=M_{U^{\pm\pm}}=0.33 v_{\chi}$.} \label{Graph3.1}
\end{figure*}

\begin{figure*}[!htp]
\centering
\includegraphics[scale=0.8]{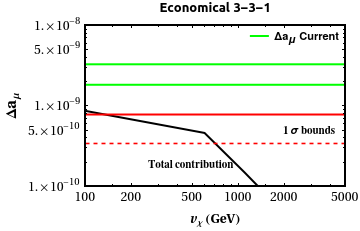}
\caption{Overall contribution to $\Delta a_\mu$ from the {\it Economical 3-3-1} model. We used $M_{Z^\prime}=0.395 v_{\chi}$,$M_{W^\prime}=M_{U^{\pm\pm}}=0.33 v_{\chi}$.} \label{Graph4}
\end{figure*}

\begin{figure*}[!ht]
\centering
\includegraphics[scale=0.7]{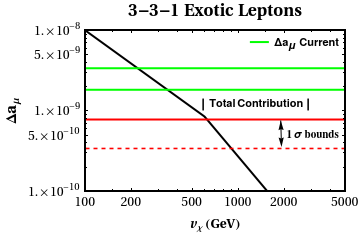}
\includegraphics[scale=0.7]{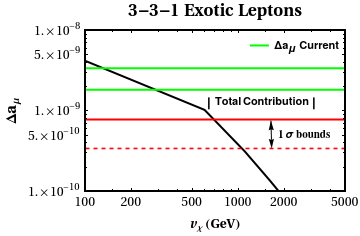}
\caption{The overall $\Delta a_\mu$ with contribution from the {\it 3-3-1 model with exotic leptons} assuming $m_N =10$~GeV, $m_E =150$~GeV (left-panel) and $m_N=100$~GeV, $m_E=500$~GeV (right-panel). We used $M_{Z^\prime}=0.55 v_{\chi}$, $M_{K^\prime}=M_{K^0}=0.46 v_{\chi}$.} 
\label{Graph5}
\end{figure*}


\subsection{3-3-1 r.h.n}

In FIG.\ref{Graph2} we show $\Delta a_\mu$ prediction from the  {\it 3-3-1 r.h.n} model. The individual contributions are shown in the {\it Appendix}. The $W^\prime$ and $Z^\prime$ gauge bosons give rise to the largest corrections to $a_\mu$. As we explained earlier we can express each correction to $a_\mu$ in terms of the scale of symmetry breaking as the particle masses depend on it. This feature allowed us to plot $\Delta a_\mu$ as a function of $v_{\chi}$ in FIG.\ref{Graph2}. We conclude from  FIG. \ref{Graph2} that $v_{\chi} < 200$~GeV is needed to reproduce the measured $\Delta a_\mu$. However, such a small value has been ruled out by collider searches for $Z^\prime$ bosons. Collider bounds impose $M_{Z^{\prime}} > 4$~TeV \cite{Lindner:2016bgg}, which implies that $v_{\chi} > 12$~TeV. We highlight that such lower mass bound on the $Z^\prime$ mass is not robust because it assumed that the $Z^\prime$ field decays only into charged leptons, which may not be true in this model as the $Z^\prime$ can also decay into exotic quarks. The presence of such decay modes certainly will weaken the LHC limit. We have implemented the model in Calchep \cite{Belyaev:2012qa} via Lanhep \cite{Semenov:2014rea} and found that indeed the $Z^\prime$ branching ratio into charged leptons can be suppressed up to $34\%$ when all three exotic quarks are sufficiently light for the $Z^\prime$ decay into. Therefore, conservatively speaking the lower mass bound on the $Z^\prime$ boson should read $2.64$~TeV, implying that $v_{\chi} > 6.68$~TeV. Hence, regardless of the inclusion of new exotic decays, collider bounds still forbid the  {\it 3-3-1 r.h.n} model to provide a solution to the muon anomalous magnetic moment. That said, we obtain lower mass bounds requirement the  {\it 3-3-1 r.h.n} contribution to $a_\mu$ to be within the $1\sigma$ error bars. The bounds are summarized in  Table \ref{bosonmassbounds}.  

\subsection{3-3-1 LHN}

The 3-3-1 model that features the presence of neutral leptons instead of right-handed neutrinos is known as  {\it 3-3-1 LHN}. This ingredient is sufficient to change our conclusions. It is a fact unexplored in the literature. Again we show for completeness the individual contributions in the {\it APPENDIX}. In Fig. \ref{Graph3.1} we show $\Delta a_\mu$ when the contribution from the  {\it 3-3-1 LHN} model are accounted for, assuming $M_N=1$~GeV. At first look it appears that the {\it 3-3-1 LHN} may explain the anomaly for $1\,{\rm TeV} < v_{\chi} < 2$~TeV. However, the collider bounds we discussed previously in the  {\it 3-3-1 r.h.n} model are in principle also applicable here. Thus, one could imagine that $v_{\chi}$ should be also larger than $12$~TeV to be consistent with current LHC bounds on the $Z^\prime$~ mass. As aforementioned, such collider bound can be weakened by the presence of $Z^\prime$ decays into exotic quarks. In this model, there are additional decay models in N's pairs. We have implemented the model in Calchep \cite{Belyaev:2012qa} via Lanhep \cite{Semenov:2014rea} and assessed the impact of these new decay in the $Z^\prime$ branching ratio into charged leptons. We concluded that branching ratio into charged leptons may be diminished up to 60\%, when also three exotic quarks and three neutral leptons, N, are light enough for the $Z^\prime$ to decay into. Conservatively speaking, it means that the lower mass bound should also be weakened by 50\%, which implies that $M_{Z^\prime}>2$~TeVs and that $v_{\chi} > 5.06$~TeV. Thus, even considering new exotic decays, the {\it 3-3-1 LHN} is not capable of addressing the muon anomaly because the scale of symmetry breaking is too small to be consistent with LHC limits. The conclusion would not change if we had adopted different values for $M_N$. Enforcing the $\Delta a_\mu$ to be smaller than $78\times 10^{-11}$ and $34 \times 10^{-11}$ we get $v_{\chi} > 407$~GeV and $v_{\chi}>722$~GeV that yield $M_{Z^\prime} >160$~GeV and $M_{Z^\prime} >  285$~GeV, respectively. A similar logic applies to the $W^\prime$ boson. Our lower mass bounds are shown in Table \ref{bosonmassbounds} even for the case when $M_N=100$~GeV.

\subsection{Economical 3-3-1}
The total correction to $a_\mu$ from the  {\it Economical 3-3-1} model is displayed in FIG. \ref{Graph4}. This model is rather similar to the  {\it 3-3-1 r.h.n} model. Thus it cannot accommodate the $a_\mu$ anomaly either. The main difference appear in the mass of the gauge bosons which have a different relation with $v_{\chi}$. This results into different lower mass bounds summarized in  Table \ref{bosonmassbounds}. Looking at  Table \ref{bosonmassbounds} we can easily conclude that the constraints derived from $a_\mu$ on this model are quite weak compared to those stemming from collider searches.

\begin{table*}[!htb]
    \begin{tabular}{|c||c|c|c|c|}
        \hline
 Model      & LHC-13TeV       & g-2 current    & g-2 projected     \\ \hline \hline
Minimal 3-3-1 & $M_{Z^\prime} >3.7$~TeV \cite{Nepomuceno:2019eaz} & $M_{Z^\prime}>434.5$~GeV   &$M_{Z^\prime}>632$~GeV      \\
            & $M_{W^\prime} >3.2 $~TeV \cite{Nepomuceno:2019eaz} &  $M_{W^\prime} > 646$~GeV   & $M_{W^\prime} >996.1$~GeV    \\\hline
3-3-1 r.h.n     & $^* M_{Z^\prime} > 2.64$~TeV \cite{Lindner:2016bgg} & $M_{Z^\prime}>158$~GeV   & $M_{Z^\prime}>276.5$~GeV      \\
            & ------- & $M_{W^\prime} > 133$~GeV & $M_{W^\prime} > 239$~GeV    \\\hline
3-3-1 LHN & $^* M_{Z^\prime} > 2$~TeV \cite{Lindner:2016bgg}& $M_{Z^\prime}>160$~GeV   &$M_{Z^\prime}>285$~GeV      \\

for $M_N=1$~GeV &----  & $M_{W^\prime} >134.3$~GeV &$M_{W^\prime} >238.3$~GeV     \\\hline
 3-3-1 LHN& $^* M_{Z^\prime} > 2$~TeV \cite{Lindner:2016bgg}& $M_{Z^\prime}>136.7$~GeV   &$M_{Z^\prime}>276.5$~GeV      \\
 for $M_N=100$~GeV           & ------  & $M_{W^\prime} >114.2$~GeV &$M_{W^\prime} >231$~GeV     \\\hline
Economical  & $^* M_{Z^\prime} >2.64 $~TeV \cite{Lindner:2016bgg}  & $M_{Z^\prime}>59.3$~GeV    &$M_{Z^\prime}>271.4$~GeV       \\
   3-3-1      & ------ & $M_{W^\prime} >49.5$~GeV  &$M_{W^\prime} >226.7$~GeV      \\ \hline
3-3-1 exotic leptons & $^* M_{Z^\prime} > 2.91$~TeV \cite{Salazar:2015gxa}  & $M_{Z^\prime}>429$~GeV   &$M_{Z^\prime}>693$~GeV       \\ 
   for $m_N=10$~GeV, $m_E=150$~GeV  & ------  & $M_{W^\prime} >359$~GeV  &$M_{W^\prime} >579.6$~GeV      \\\hline 
3-3-1  exotic leptons & $^* M_{Z^\prime} > 2.91$~TeV \cite{Salazar:2015gxa}  & $M_{Z^\prime}>369$~GeV   &$M_{Z^\prime}>600$~GeV       \\ 
 for $m_N =10$~GeV, $m_E =150$~GeV   & ------  & $M_{W^\prime} >309.1$~GeV  &$M_{W^\prime} >501.4$~GeV      \\\hline
    \end{tabular}
    \caption{Summary of the lower bounds based on our calculations. For comparison we include the LHC bounds at 13~TeV center-of-mass energy. These LHC bounds are based on either $36 fb^{-1}$ or $139fb^{-1}$ of data. The lower mass bounds on the $Z^\prime$ and $W^\prime$ bosons from the  {\it 3-3-1 r.h.n} model are also applicable to the {\it Economical 3-3-1} model as they have the same interactions. We emphasize that the limits quoted in \cite{Lindner:2016bgg} neglected $Z^\prime$ decays into exotic quarks and heavy leptons which can weaken the lower mass bounds up to 50\%. The same is true for the bounds obtained in \cite{Salazar:2015gxa}. The effect of $Z^\prime$ exotic decays were considered in the {\it Minimal 3-3-1} model in \cite{Nepomuceno:2019eaz}. Conversely, the lower mass constraints we found are robust.}
    \label{bosonmassbounds}
\end{table*}

\subsection{3-3-1 with exotic leptons}

The  3-3-1 model with exotic leptons features gauge bosons, $K^{\pm}$ and $K^0$, that induce the largest corrections to $a_\mu$. The interactions that these bosons experience are  absent in the previous models, because they couple to exotic charged leptons, E, which can be heavy. The values obtained for $\Delta a_\mu$ are shown in FIGs. \ref{Graph5} for $M_N=10$~GeV, $M_E=150$~GeV (left-panel) and  $M_N=100$~GeV, $M_E=500$~GeV (right-panel). We considered these two cases to assess the dependence of our finding on the  masses of these particles. We remind that the individual contributions are also exhibited in the {\it Appendix}. The total contribution to $a_\mu$ is rather small because there more than one heavy field running in the loop.  Consequently, the scale of symmetry breaking needed to reproduce the measured $\Delta a_\mu$ value is too small to be consistent with collider bounds. Therefore, once again, we can simply derive $1\sigma$ lower bound on scale of symmetry breaking which can be translated into lower mass limits on the gauge boson masses as shown in  Table \ref{bosonmassbounds}.   

\section{Difference between 3-3-1 Models}

We highlight that the gauge bosons drive the contributions to g-2. In simplified model constructions the masses of the gauge bosons can be made independent quantities. Here, this does not occur. First, as a result of the $SU(3)_L$ nature, we have multiple gauge bosons, and the pattern of symmetry breaking ties the gauge bosons masses to a single parameter, $v_{\chi}$. For this reason our findings are expressed in terms of $v_{\chi}$. One may wonder about the impact of scalar fields in our numerical calculations. As aforementioned, they yield negligible contributions, regardless of the Yukawa couplings assumed. Concerning the masses of the exotic fermions, these do impact the overall corrections to g-2, but we remind the reader that these contributions are still mediated by gauge bosons. We have assessed how the masses of the exotic fermions impact our findings, for instance in FIGs.4-6. 

In summary, as far as g-2 is concerned, one can clearly see the difference between the 3-3-1 models by comparing the FIGS.2-6. The {\it Minimal 3-3-1} model gives rise to a much larger correction to g-2 in comparison with the {\it 3-3-1 r.h.n} model as a result of the doubly charged gauge boson. The {\it 3-3-1 LHN} model differs from the {\it 3-3-1 r.h.n} prediction mainly due to the presence of a neutral fermion whose masses lies in the $1-100$~GeV mass range. Such fermion is absent in the {\it 3-3-1 r.h.n} model. The {\it Economical 3-3-1} models which is very similar to the {\it 3-3-1 r.h.n} model still in the end of day yield a different overall contribution to g-2 because the masses of the gauge bosons have a slight different dependence with $v_{\chi}$. The {\it 3-3-1 model with exotic leptons} despite the distinct mass spectrum may give rise to an overall correction to g-2 which is similar to one stemming from the {\it 3-3-1 LHN} depending on the values adopted for the masses of the exotic fermions. Anyway, our finding show that all these 3-3-1 models yield different corrections to g-2 and if the g-2 anomaly is confirmed they must be extended.

\section{Discussions}

\begin{figure}[!htp]
    \centering
    \includegraphics[scale=0.67]{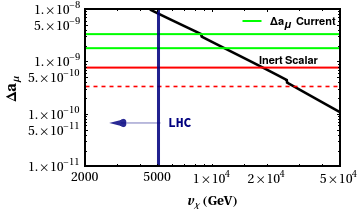}
    \caption{Overall contribution of the {\it 3-3-1 LHN} model augmented by an inert scalar triplet $\phi$. The LHC bound is represented by a vertical line which excluded the region of parameter space to the left. One can notice that this extended version of the {\it 3-3-1 LHN} model now successfully accommodate the $a_\mu$ anomaly for $v_{\chi}\sim 10$~TeV, while being consistent with LHC constraint. }
    \label{fig:ultima}
\end{figure}

We have shown that all five models studied cannot accommodate the muon anomalous magnetic moment anomaly in agreement with existing bounds. This finding is timely important because we expect a new measurement of the muon anomalous magnetic moment. Moreover, there is an ongoing discussion concerning the significance of the signal in light of large hadronic uncertainties. If the g-2 anomaly turns out to be just a statistical fluctuation our results represent a lower bound on the energy scale at which the 3-3-1 symmetry should be broken. Our conclusion holds even if one departs from the $v_{\eta}=v_{\rho}$ assumption adopted throughout. These choices for $v_{\eta}$ and $v_{\rho}$ affect more the scalar contributions which are already suppressed by the muon mass. On the other hand, if the g-2 anomaly is confirmed, one ought to think of ways of extending such models to address $a_\mu$ while being consistent with collider bounds. The addition of an inert scalar triplet under $SU(3)_L$ \cite{CarcamoHernandez:2020pxw}, or inert singlet scalar, or vector-like leptons, etc., represent viable possibilities \cite{Queiroz:2014zfa}. However, each of these avenues constitute a new model, with their own phenomenological implications that we plan to explore in the near future. For concreteness, we have added an inert scalar triplet, $\phi$, into the  {\it 3-3-1 LHN} model and computed the overall contribution to $a_\mu$. Such scalar triplet which gets a mass from the quartic coupling in the scalar potential, $\lambda \phi^\dagger \phi \chi^\dagger \chi$, after the scalar triplet $\chi$ acquires a $vev$. This  mass goes as $M_\phi \sim \lambda v_{\chi}$.  This inert triplet scalar allow us to include $\mathcal{L} \supset y_{ab} \bar{f_a} \phi e_{bR}$. Taking $y_{22}$ equal to unit we get FIG. \ref{fig:ultima}. There we display the total contribution of the  3-3-1 LHN model augmented by this inert scalar triplet. We can successfully explain the muon anomalous magnetic moment anomaly for $v_{\chi} \sim 10$~TeV, while being consistent with LHC constraint that rules out the region with $v_{\chi} < 5.06$~TeV. Therefore, we have conclusively presented a solution to the muon anomalous magnetic moment in the context of 3-3-1 models.

\section{Conclusions}

We have revisited the contribution to the muon anomalous magnetic moment stemming from five different models based on the $SU(3)_C\times SU(3)_L\times U(1)_N$ gauge symmetry. We have assessed the impact of changing the masses of the exotic leptons present in such models and shown that our quantitative conclusions do change depending on the value assumed for their masses. A fact unexplored before. Moreover, we corrected previous estimations in the literature such as the contribution stemming from the doubly charged gauge boson, as it does not have a vectorial current to muons. Moreover, we have drawn our conclusions in perspective with collider bounds and concluded that none of the five models investigated here are capable of accommodating the anomaly. Consequently, we derived robust and complementary $1\sigma$ lower mass bounds on the masses of the new gauge bosons, namely the $Z^\prime$ and $W^\prime$ bosons. In summary, if the anomaly observed in the muon anomalous magnetic moment is confirmed by the g-2 experiment at FERMILAB these models must be extended. For concreteness, we presented a plausible extension to the  {\it 3-3-1 LHN} model, which features an inert scalar triplet. This extension can accommodate the anomaly for $v_{\chi} \sim 10$~TeV, while being consistent with LHC limits. We make our {\it Mathematica numerical codes} available at \cite{Mathematicacode} to allow the reader to double check our findings and apply our tool to other studies of the muon anomalous magnetic moment.

\section*{Acknowledgement}

The authors are grateful to Antonio Santos for discussions. SK acknowledges support from ANID-Chile Fondecyt No. 1190845 and 
ANID-Chile PIA/APOYO AFB180002. ASJ and YSV acknowledge support from CAPES. CASP is supported by the CNPq research grants No. 304423/2017-3. FSQ thanks CNPq grants 303817/2018-6 and 421952/2018-0, and ICTP-SAIFR FAPESP grant 2016/01343-7 for the financial support. This work was supported by the
Serrapilheira Institute (grant number Serra-1912-31613). We thank the High Performance Computing Center (NPAD) at UFRN for providing computational resources.

\section{Appendix}

\begin{figure*}[!htp]
    \centering
\includegraphics[scale=0.7]{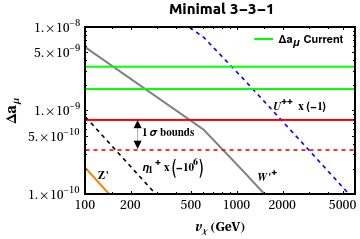}
\includegraphics[scale=0.7]{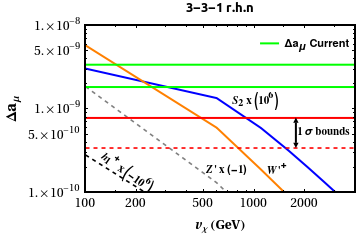}
    \caption{Overall $\Delta a_\mu$ taking into account each individual contribution from the {\it Minimal 3-3-1} model and {\it 3-3-1 r.h.n.} }
    \label{figind1}
\end{figure*}

Here, in Figs.~\ref{figind1}-\ref{figind4} we show for completeness  the plots of individual contributions to $a_\mu$ of the particles introduced in the main text.

\begin{figure*}[!htp]
    \centering
    \includegraphics[scale=0.8]{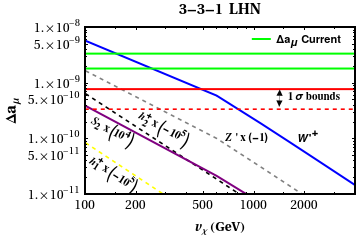}
    \caption{Overall $\Delta a_\mu$ taking into account each individual contribution from the {\it 3-3-1 LHN} model for $m_N=1$~GeV}
\end{figure*}

\begin{figure*}[!htp] 
\centering
\includegraphics[scale=0.7]{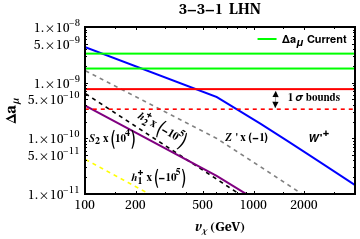}
\includegraphics[scale=0.7]{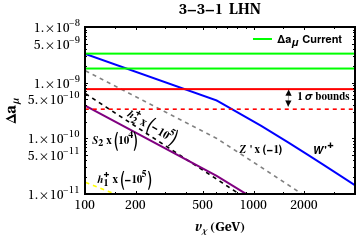}
    \caption{Overall $\Delta a_\mu$ taking into account each individual contribution from the {\it 3-3-1 LHN} model for $m_N=100$~GeV (left-panel) and $m_N=240$~GeV (right-panel).}
    \label{figind2}
\end{figure*}

\begin{figure*}[!htp]
    \centering
\includegraphics[scale=0.7]{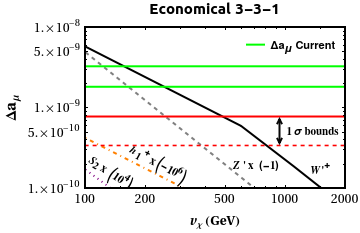}
    \caption{Overall $\Delta a_\mu$ taking into account each individual contribution from the {\it Economical 3-3-1} model.}
    \label{figind3}
\end{figure*}

\begin{figure*}[!htp]
    \centering
\includegraphics[scale=0.7]{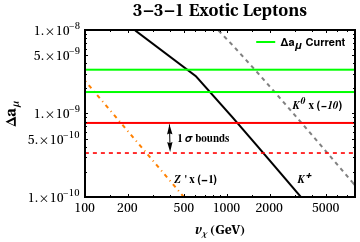}
\includegraphics[scale=0.7]{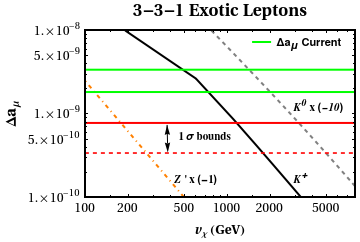}
    \caption{Overall $\Delta a_\mu$ taking into account each individual contribution from the {\it 3-3-1 with exotic leptons}, for $m_N =10$GeV, $m_E =150$~GeV (left-panel) and $m_N=100$GeV, $m_E=500$~GeV (right-panel).}
    \label{figind4}
\end{figure*}

\nocite{*}
\bibliography{ref}

\end{document}